\documentstyle[11pt,IAUS215,twoside,epsf]{article}

\newcommand{\Msun}{{\rm M}_{\scriptscriptstyle \odot}}
\def\ltaprx {\lower .1ex\hbox{\rlap{\raise .6ex\hbox{\hskip .3ex
        {\ifmmode{\scriptscriptstyle <}\else 
                {$\scriptscriptstyle <$}\fi}}}
        \kern -.4ex{\ifmmode{\scriptscriptstyle \sim}\else 
                {$\scriptscriptstyle\sim$}\fi}}}
\def\gtaprx {\lower .1ex\hbox{\rlap{\raise .6ex\hbox{\hskip .3ex
        {\ifmmode{\scriptscriptstyle >}\else 
                {$\scriptscriptstyle >$}\fi}}}
        \kern -.4ex{\ifmmode{\scriptscriptstyle \sim}\else 
                {$\scriptscriptstyle\sim$}\fi}}}

\def\edcomment#1{\iffalse\marginpar{\raggedright\sl#1\/}\else\relax\fi}
\reversemarginpar

\begin{document}
\vspace*{1cm}
\title{Supernovae, Gamma-Ray Bursts, and Stellar Rotation}
 \author{S. E. Woosley}
\affil{Department of Astronomy and Astrophysics, UCSC, Santa Cruz CA
  95064, USA}
\author{A. Heger}
\affil{Department of Astronomy and Astrophysics, Enrico Fermi
  Institute, The University of Chicago, 5640 S. Ellis Avenue, Chicago
  IL 60637, USA}

\begin{abstract}
One of the most dramatic possible consequences of stellar rotation is
its influence on stellar death, particularly of massive stars. If the
angular momentum of the iron core when it collapses is such as to
produce a neutron star with a period of 5 ms or less, rotation will
have important consequences for the supernova explosion
mechanism. Still shorter periods, corresponding to a neutron star
rotating at break up, are required for the progenitors of gamma-ray
bursts (GRBs). Current stellar models, while providing an excess of
angular momentum to pulsars, still fall short of what is needed to
make GRBs. The possibility of slowing young neutron stars in ordinary
supernovae by a combination of neutrino-powered winds and the
propeller mechanism is discussed. The fall back of slowly moving
ejecta during the first day of the supernova may be critical. GRBs, on
the other hand, probably require stellar mergers for their production
and perhaps less efficient mass loss and magnetic torques than
estimated thus far.
\end{abstract}

\section{Introduction}

For seventy years scientists have debated the physics behind the
explosion of massive stars as supernovae (Baade \& Zwicky 1934;
Colgate \& White 1966) and the possible role of rotation in these
events (Hoyle 1946; Fowler \& Hoyle 1964; LeBlanc \& Wilson 1970;
Ostriker \& Gunn 1971). During the last five years, the discovery of
supernovae showing evidence for gross asymmetry and anomalously large
energy (e.g., Kawabata et al. 2002) coupled with a growing realization
that GRBs require rapidly rotating massive stars for their
progenitors, has rekindled interest in this historic controversy.

Here we briefly review ($\S$2) the currently favored models for GRBs.
The starting point for each is a rapidly rotating, massive (M $\gtaprx
\, 10 \, \Msun$) Wolf-Rayet star (WR-star) whose central iron core has
become unstable. The amount of angular momentum required for the
material just outside the core ($j \gtaprx 10^{16}$ cm$^2$ s$^{-1}$)
is far in excess of what is observed in young pulsars and also greater
than what is predicted by current stellar models (Heger, Woosley, \&
Spruit 2003; see also Heger et al. this volume). In fact, the models
predict angular momenta in presupernova stars in between what is
needed for GRBs and pulsars - too slow for the former, too fast for
the latter ($\S$3). We are thus led to consider scenarios in which
excess angular momentum exists in a small fraction of WR-stars near
death, as well as mechanisms for {\sl removing} angular momentum from
neutron stars produced by ordinary supernovae ($\S$4).  Given that
current estimates of gravitational radiation by the $r-$mode
instability are orders of magnitude too small, a promising possibility
is braking the rotation of young neutron stars by the ``propeller
mechanism''. This happens when a strongly magnetic neutron star is
spun down by interacting with slow moving matter that fails to achieve
ejection in the supernova. As this material falls back on the spinning
neutron star, it is spun up and ejected, carrying with it excess
angular momentum.

We conclude with some speculations ($\S$5) on how the diverse needs
for slowly rotating pulsars and rapidly rotating GRB progenitors might
be achieved.

\section{Models for Gamma-Ray Bursts}

Recent developments in the study of GRBs have been reviewed by
Meszaros (2002) and we will only mention a few relevant to the present
discussion. It is now acknowledged that the most common variety of
GRBs, the so called ``long-soft'' bursts, originate at cosmological
distances with an average redshift of about 1.3 (there may be
observational bias in this number and the true average could be
greater). Each burst is beamed to a small fraction of the sky with a
typical opening angle of $\sim$5 degrees; thus for every burst we see,
there are approximately 300 that we do not see. Correcting for this
beaming and making a reasonable estimate for the efficiency of
producing gamma-rays, the kinetic energy of the GRB producing jets is
$\sim 3 \times 10^{51}$ erg (Frail et al 2001). Approximately two
dozen events cluster within an order of magnitude of this this
value. It is interesting that this is close to the kinetic energy of a
typical supernova ($1 \times 10^{51}$ erg), but, on the other hand,
ordinary supernovae do not collimate their ejecta into jets and do not
concentrate a large fraction of their energy into highly relativistic
motion. It is estimated, on the basis of variability arguments and
self-opacity considerations, that the Lorentz factor in these jets is
typically in excess of $\Gamma \sim$ 300, making them by far the most
rapidly moving bulk matter in the universe. The event rate in the
observable universe for GRBs, both those beamed at us and away, is
about 1000 per day.

Observations of GRB afterglows, where they have been seen, show
the bursts come from star-forming regions in distant galaxies. There
have also been reports of approximately five Type I supernovae
coincident in location and onset with GRBs. The most striking and well
observed of these was SN 1998bw, in coincidence with GRB 980425. This
supernova, a Type Ic, had very unusual properties including anomalously
high velocities, high energy, asymmetry, and evidence for relativistic
motion (inferred from its radio afterglow).

Taken together, the evidence suggests that at least this common
subclass of GRBs is produced by the death of massive stars. The
energies involved are similar to supernovae, but somehow a large
fraction of the energy has been channeled into relativistic, bi-polar
outflows. We see a burst only when we look directly down the axis of
these outflows. Finally, since the bursts typically last $\sim$20 s,
the star producing the GRB cannot be a giant star (neither blue nor
red) because a jet lasting 20 s or less would lose its energy before
escaping from the star. It would also be unrealistic for a jet that
lasted much longer, say the $\sim$1000 seconds required for light to
cross a red supergiant, to die abruptly 10 - 20 s after escaping.  The
progenitor has apparently lost its envelope.

Within this context several models have been proposed. Each involves
the death of a WR-star initiated by iron-core collapse, and in
each, the parent star is exploded by jets that then travel very far
outside the stellar surface before making the actual GRB. As we shall
see, each model, also invokes rapid rotation of the parent star.

\subsection{Supranovae}

Vietri \& Stella (1998, 1999) have suggested that GRBs result from the
delayed implosion of rapidly rotating neutron stars to black
holes. The neutron star forms in a traditional (neutrino-powered)
supernova, but is ``supramassive'' in the sense that without rotation,
it would collapse. With rapid rotation, however, collapse is delayed
until angular momentum is lost by gravitational radiation and magnetic
field torques.  Vietri and Stella assume that the usual pulsar formula
holds and, for a field of 10$^{12}$ G, a delay of order years
(depending on the field and mass) is expected. When the centrifugal
support becomes sufficiently weak, the star experiences a period of
runaway deformation and gravitational radiation before collapsing into
a black hole. It is assumed that $\sim0.1 \, \Msun$ is left behind in a
disk which accretes and powers the burst explosion.

As a GRB model, the supranova has several advantages. It predicts an
association of GRBs with massive stars and supernovae - as do all the
models discussed here. Moreover it produces a large amount of material
enriched in heavy elements located sufficiently far from the GRB as
not to obscure it. The irradiation of this material by the burst or
afterglow can produce x-ray emission lines as have been reported in
several bursts (e.g., Piro et al. 1999, 2000).  However, the supranova
model also has some difficulties.  It may also take fine tuning to
produce a GRB days to years after the neutron star is born. Shapiro
(2000) has shown that neutron stars requiring differential rotation
for their support will collapse in only a few minutes.  The
requirement of rigid rotation reduces the range of masses that can be
supported by rotation to, at most, $\sim$20\% above the non-rotating
limit. The angular momentum required is essentially that of a neutron
star born rotating at break up.

\subsection{Millisecond magnetars}

Another model, championed most recently by Wheeler et al. (2000,
2002), is the ``super-magnetar'' model (see also Usov 1992). As usual,
the iron core of a massive star collapses to a neutron star, but for
whatever reasons, unusually high angular momentum perhaps, the neutron
star acquires, at birth, an extremely powerful magnetic field,
$10^{15}$ - 10$^{17}$ G. If the neutron star additionally rotates
with a period of $\sim 1$ ms, up to 10$^{52}$ erg in rotational energy
can be extracted on a GRB time scale by a variation of the pulsar
mechanism. This model has the attractive features of being associated
with massive stars, making a supernova as well as a GRB, and utilizing
an object, the magnetar, that is implicated in other phenomena - soft
gamma-ray repeaters and anomalous x-ray pulsars. It has the
unattractive feature of invoking the magnetar fully formed in the
middle of a star in the process of collapsing without consideration of
the effects of neutrinos or rapid accretion. The star does not have
time to develop a deformed geometry or disk that might help to
collimate jets. To break the symmetry, Wheeler et al. invoke the
operation of a prior LeBlanc-Wilson (1970) jet to ``weaken'' the
confinement of the radiation bubble along the rotational
axis. Numerical models to give substance to this scenario are needed
(see also Wheeler, Meier, \& Wilson 2002).

\subsection{Collapsars}

This is currently the leading model for GRBs, in part because of the
detailed calculations that have been done to support the scenario.  A
collapsar is a rotating massive star whose central core collapses to a
black hole surrounded by an accretion disk (Woosley 1993; MacFadyen \&
Woosley 1999). Accretion of at least a solar mass through this disk
produces outflows that are further collimated by passage through the
stellar mantle. These flows attain high Lorentz factor as they emerge
from the stellar surface and, after traversing many stellar radii,
produce a GRB and its afterglows by internal and external shocks. The
passage of the jet through the star also gives a very asymmetric
supernova of order 10$^{51}$ erg.

There are three ways to make a collapsar and each is likely
to have different observational characteristics.

\begin{itemize}

\item 
{A standard (Type I) collapsar (MacFadyen \& Woosley 1999) is one
where the black hole forms promptly in a helium core of approximately
15 to 40 $\Msun$. There never is a successful outgoing shock after the
iron core first collapses. A massive, hot proto-neutron star briefly
forms and radiates neutrinos, but the neutrino flux is inadequate to
halt the accretion.  Such an occurrence seems likely in helium cores
of mass over $\sim10 \, \Msun$ because of their large binding energy
and the rapid accretion that characterizes the first second after core
collapse.}

\item
{A variation on this theme is the ``Type II collapsar'' wherein the
black hole forms after some delay - typically a minute to an hour,
owing to the fallback of material that initially moves outwards, but
fails to achieve escape velocity (MacFadyen Woosley, \& Heger
2001). Such an occurrence is again favored by massive helium
cores. Unfortunately the long time scale associated with the fall back
may be, on the average, too long for typical long, soft bursts. Their
accretion disks are also not hot enough to be neutrino dominated and
this may affect the accretion efficiency (Narayan, Piran, \& Kumar
2001) and therefore the energy available to make jets.}

\item
{A third variety of collapsar occurs for extremely massive
metal-deficient stars (above $\sim$300 $\Msun$) that probably existed
only in the early universe (Fryer, Woosley, \& Heger 2001). For
non-rotating stars with helium core masses above 133 $\Msun$ (main
sequence mass 260 $\Msun$), it is known that a black hole forms after
the pair instability is encountered. It is widely suspected that such
massive stars existed in abundance in the first generation after the
Big Bang at red shifts $\sim$5 - 20. For {\sl rotating} stars the mass
limit for black hole formation will be raised. The black hole that
forms here, about 100 $\Msun$, is more massive, than the several
$\Msun$ characteristic of Type I and II collapsars, but the accretion
rate is also much higher, $\sim$10 $\Msun$ s$^{-1}$, and the energy
released may also be much greater. The time scale is also much
longer.}

\end{itemize}

Of these possibilities, the collapsar Type I is currently favored and
most frequently discussed, in part because of its shorter
characteristic time scale.

\section{GRB Progenitors}

Using the same formalism and code as discussed by Heger et al. (these
proceedings), a 15 $\Msun$ {\sl helium} core has been evolved from the
onset of central helium burning until iron core collapse. Estimates of
the dominant magnetic and non-magnetic modes of angular momentum
transport were included. The assumed mass loss rate was taken from
Wellstein \& Langer (1999) reduced by 3 to account for clumping
(Hamann \& Koesterke 1998). We take a metallicity of 10\% solar and
further decrease the mass loss by the square root of this
number. Altogether the mass loss rate was
\begin{equation}
\log \, (\frac{\dot M}{\Msun \ {\rm yr^{-1}}}) \ = \ -10.95 \ + \ 1.5 \,
\log \, (\frac{L}{L_{\scriptscriptstyle \odot}}).
\end{equation}
This is equal to $1.7 \times 10^{-5} \ \Msun$ yr$^{-1}$ when the star
dies, a value some might regard as low for an 8.6 $\Msun$ WR-star but
at least an order of magnitude greater than inferred for SN
2002ap (Berger, Kulkarni, \& Chevalier 2002) and SN 1998bw (Li \&
Chevalier 1999), two well studied massive Type Ic supernovae. The
initial model was assumed to be rigidly rotating with a surface
angular speed corresponding to a fraction of break up (typically 10\%
or 30\%). A sampling of results is given in Table 1 and in the
figures. For other models and more extensive discussion see Heger,
Woosley, \& Spruit (2003).

The first three columns in Table 1 define the initial model and
physics employed (magnetic torques, amount of rotation expressed in
terms of a per-centage of Keplerian at the surface, and mass loss
by winds).  Next is the period a pulsar would have if it formed in
this star, then the non-dimensional spin parameter, $a$, a black hole
would acquire, if all the angular momentum below the mass coordinate
indicated were to go into the black hole of that mass (formal values
in excess of 1 are shown in brackets solely to give a measure for the
angular momentum available in the model). In the last column we show
the mass ranges in which the equatorial mass could form a
centrifugally supported accretion disk around a central compact
object.  The dashed lines in Fig. 2 indicate the specific angular
momentum needed for a rotationally supported disk at the last stable
orbit of a Schwarzschild or extreme Kerr (a = 1) black hole.

As the figures and table show, mass loss and magnetic torques are
detrimental to any GRB model that invokes hyper-critical rotation of
the collapsed remnant. This is true even for initial rotation rates
unlikely to be achieved in single stars (e.g., 30\% critical at the
surface). Most of the angular momentum is lost during helium burning
and very little in the advanced burning stages.

\vskip 0.2 in
\centerline {TABLE 1}
\begin{table}
\begin{tabular*}{\columnwidth}{@{}c@{\extracolsep{\fill}}c@{\extracolsep{\fill}}c|@{}c|@{}c@{\extracolsep{\fill}}c@{\extracolsep{\fill}}c@{\extracolsep{\fill}}c|@{}c@{}}
B & \% K & $\dot M$ & P(ms) & a(2) & a(2.5) & a(3) & a(4) & Mass range
for making disk
\\
\hline
 -- & 10 &  -- & 0.09 & (2.5) & (3.4) & (3.8) & (3.7) & $ 0-15$ \\
 -- & 10 & yes & 0.23 & (1.1) &  0.90 &  0.98 & (1.2) & $ 0- 2.5$, $ 2.7- 8.6$ \\
 -- & 30 &  -- & 0.06 & (4.1) & (4.8) & (5.6) & (5.9) & $ 0-15$ \\
 -- & 30 & yes & 0.18 & (1.7) & (1.3) & (1.3) & (1.7) & $ 0- 8.4$ \\
\hline           
yes & 10 &  -- & 0.45 &  0.58 &  0.76 &  0.80 &  0.88 & $ 2.4- 2.8$, $ 5.7-15$ \\
yes & 10 & yes &  3.4 &  0.08 &  0.09 &  0.11 &  0.10 & $ 8.5- 8.8$ \\
yes & 30 &  -- & 0.26 & (1.2) & (1.4) & (1.4) & (1.5) & $ 0-15$ \\
yes & 30 & yes &  1.9 &  0.16 &  0.14 &  0.16 &  0.17 & $ 7.4- 8.8$ \\
\hline           
\end{tabular*}
\end{table}

\begin{figure}
\plotone{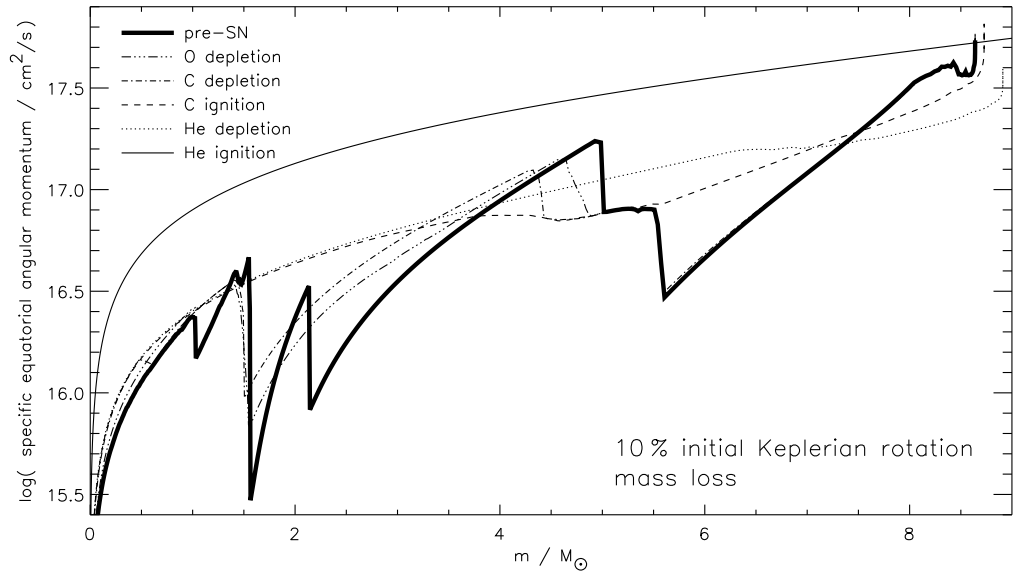}
\caption{Angular momentum in the equatorial plane as a function of the
interior mass coordinate, $m$, at different evolutionary stages for a
15 $\Msun$ helium star. Mass loss was included, but no magnetic
torques.  The specific angular momentum is shown during different
evolutionary stages and for the presupernova star. The peak-like
structure in $j(m)$ at late times is indicative of rigidly rotating
convective shells. This model has adequate rotation at death to
produce a collapsar (line 2; Table 1).
\label{fig:b3jj}}
\end{figure}

\section{Slowing the Rotation Rate of Young Neutron Stars}

While the rotation rates in Fig. 2 and Table 1 are too slow for GRB
progenitors, they are much too fast for ordinary pulsars. This remains
true even for the more moderate rotation rates coming from red
supergiant progenitors and discussed by Heger et al. (this volume).  A
young neutron star can be braked by neutrino emission, by
gravitational radiation, and by electromagnetic torques. The first two
have the virtue of carrying away angular momentum and energy in a form
that, at least for now, is invisible. In principle, core collapse
could occur in the presence of rotation so large as to be dynamically
important, or even to make a collapsar, but the excess angular
momentum could be radiated away before any measurements of the pulsar
period become possible.

\begin{figure}
\plotone{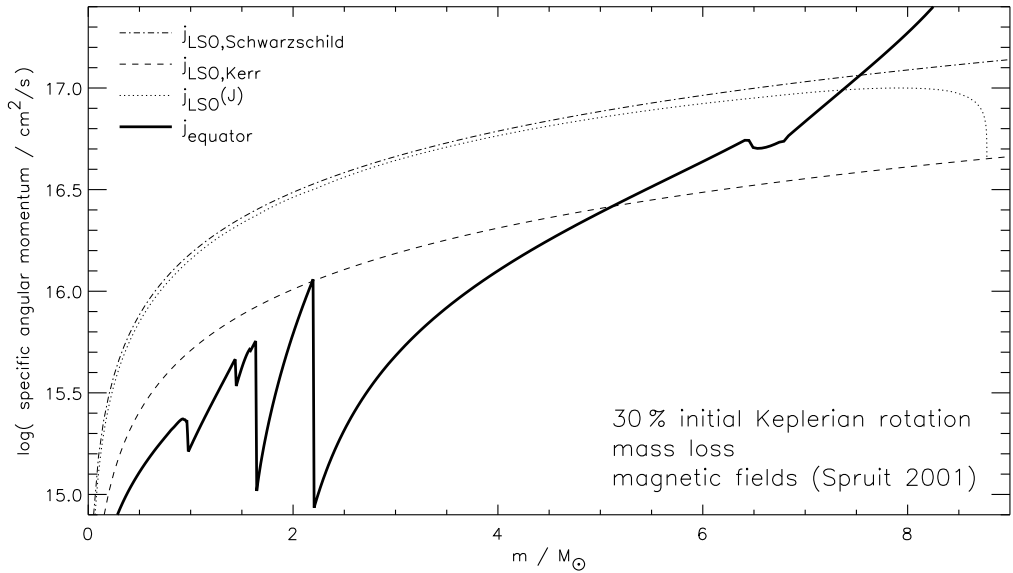}
\caption{Same as Fig. 1, but with an initial rotation rate of 30\,\%
Keplerian rotation and angular momentum transport by magnetic fields
according to Spruit (2002) included. Dashed lines show the necessary
angular momentum to give a rotationally supported disk for a
Schwarzschild black hole (upper curve) or a Kerr Black hole (lower
curve). Only the outermost layers of the core satisfy either condition
and the model is not a viable collapsar progenitor (bottom line, Table
1). However, less mass loss or less efficient magnetic torques could
change this conclusion.}
\label{fig:g3j}
\end{figure}

Neutrinos carry away angular momentum because their last interaction
occurs at the surface of the neutron star where the specific angular
momentum is high. The ratio of angular momentum to effective mass
(E/c$^2$) is greater than the average for the neutron star and
neutrino loss results in the loss of specific angular momentum in the
remaining star. Unfortunately, such losses are limited by the small
fraction of the neutron star's mass emitted as neutrinos ($\sim$20\%).
Calculations by Janka (2002, private communication) show that the
maximum lengthening of the period is 30\%.

For some time it was thought that gravitational radiation induced by
the ``$r$-mode instability'' would rapidly brake young neutron stars.
However, Arras et al (2002) have recently found that the $r$-mode waves
saturate at amplitudes far lower than obtained in previous numerical
calculations (Lindblom, Tohline, \& Vallisneri 2001) that assumed an
unrealistically large driving force. Much less gravitational radiation
occurs and the braking time for a rapidly rotating neutron star
becomes millennia rather than hours. Arras et al. calculate that
\begin{equation}
\tau_{\mathrm spin \ down} = 2000 \ {\rm yr} \ (\frac{\alpha_e}{0.1})^{-1} \
(\frac{335 \ {\rm Hz}}{\nu})^{11},
\end{equation}
with $\alpha_e$, the maximum amplitude of the instability (for which
Arras et al estimate 0.1 as an upper bound). A 3 ms pulsar will thus
take about 2000 years to substantially brake.  Kaspi \& Helfand
(2002), on the other hand, note several slowly rotating young pulsars,
in particular J1811-1925 in G11.2-0.3 with P$_{\mathrm init}$ = 62 ms and age
2000 years. It seems that pulsars are either born rotating slowly -
because their parent stars rotated slowly - or else have been braked
by electromagnetic torques. In the later case, the excess energy
should either appear as radiation or contribute to the kinetic energy
of the explosion.

In this regard the estimated rotation rates calculated for
presupernova cores by Heger et al. elsewhere in these proceedings lie
in an interesting range, too slow to be the sole source of the
supernova's energy, yet too fast for ordinary pulsars.  Such
calculations are admittedly very uncertain, but, taken at face value,
raise an interesting question. Could pulsars frequently be born
rotating with 5 to 10 ms periods, yet be slowed to values 30 to 100
ms, say, during their first year? One possibility is the pulsar
mechanism itself, but we would like to consider here two others -
magnetic stellar winds and the propeller mechanism.

\subsection{Neutrino-Powered Magnetic Stellar Winds}

During the first 10 s of its life a neutron star emits about 20\% of
its mass as neutrinos. This powerful flux of neutrinos passing through
the proto-neutron star atmosphere drives mass loss and the neutron
star loses $\sim0.01 \Msun$ (Duncan, Shapiro, \& Wasserman 1986; Qian
\& Woosley 1996; Thompson, Burrows, \& Meyer 2001). Should magnetic
fields enforce corotation on this wind out to a radius of 10 stellar
radii, appreciable angular momentum could be lost from a 1.4 $\Msun$
neutron star (since $j \sim r^2 \omega$). The issue is thus one of
magnetic field strength and its radial variation.

For a neutron star of mass $M$ and radius, $R_6$, in units of 10 km,
the mass loss rate is approximately (more accurate expressions are
given by Qian \& Woosley)
\begin{equation}
\dot M \ \approx \ 10^{-3} \ \Msun \ {\rm s^{-1}}\
\left(\frac{L_{\nu,{\rm tot}}}{10^{53} \ \rm erg \ s}\right)^{5/3} \
{R_6}^{5/3} \ \left(\frac{1.4 \ \Msun}{M}\right)^2.
\end{equation}
The field will cause corotation of this wind out
to a radius (Mestel \& Spruit 1987) where 
\begin{equation}
\rho \ (v_{\rm wind}^2 \ + \  \omega^2 r^2) \ \approx \ \frac{B^2}{4
  \pi}.
\end{equation}
Calculations that include the effect of rotation on the
neutrino-powered wind have yet to be done, but one can estimate the
radial density variation from the one-dimensional models of Qian \&
Woosley. For a mass loss rate of 10$^{-2} \ \Msun$ s$^{-1}$ they find
a density and radial speed at 100 km of 10$^8$ g cm$^{-1}$ and 1000 km
s$^{-1}$ respectively. At this time, the proto-neutron star radius is
30 km and these conditions persist for $\sim 1$ s. Later, they find,
for a mass loss rate of 10$^{-5}$ $\Msun$ s$^{-1}$ and a radius of 10
km, a density at 100 km of $\sim$10$^5$ g cm$^{-3}$ and velocity 2000
km s$^{-1}$. This lasts $\sim 10$ s. Unless $\omega$ is quite low,
$v_{\rm wind}$ is not critical. For $\omega \sim 1000$ rad s$^{-1}$
the field required to hold 10$^8$ g cm$^{-3}$ in corotation at 100 km
is $\sim 3 \times 10^{14}$ G; for 10$^5$ g cm$^{-3}$ it is
10$^{13}$ G.

To appreciably brake such a rapidly rotating neutron star, assuming $B
\sim r^{-2}$, thus requires ordered surface fields of $\sim$10$^{15}$
- 10$^{16}$ G. Similar conclusions have been reached independently
by Todd Thompson (2003). Such fields are characteristic of magnetars,
but probably not of ordinary neutron stars. On the other hand if the
rotation rate were already slow, $\omega \sim 100$ rad s$^{-1}$ ($P
\sim $ 60 ms), even a moderate field of 10$^{14}$ G could have an
appreciable effect. It should also be kept in mind that the field
strength of a neutron star when it is 1 - 10 s old could be very
different than thousands of years later when most measurements 
have been made.

\subsection{Fallback and the Propeller Mechanism} 

After the first 1000 s, the rate of accretion from fall back is given
(MacFadyen, Woosley, \& Heger 2001) by
\begin{equation}
\dot M \ \approx \ 10^{-7} \ t_5^{-5/3} \ \Msun \ {\rm s^{-1}},
\end{equation}
with the time in units of 10$^5$ s.  For the mass loss rate in 
units of $10^{26}$ g s$^{-1}$, we obtain
\begin{equation}
\dot M_{26} \ \approx \ 2 \ t_5^{-5/3} \ {\rm g \ s^{-1}}.
\end{equation}
For a dipole field with magnetic moment $\mu_{30} = B_{12} {R_6}^3$,
with $B_{12}$ the surface field in units of 10$^{12}$ G, the infalling
matter will be halted by the field at the Alfven radius (Alpar 2001),
\begin{equation}
r_{\rm A} \ = \ 6.8 \ \mu_{30}^{4/7} \ \dot M_{26}^{-2/7} \ {\rm km}.
\end{equation}
At that radius matter can be rotationally ejected, provided the
angular velocity there corresponding to co-rotation exceeds the
Keplerian orbital speed. The ejected matter carries away angular
momentum and brakes the neutron star. This is the propeller mechanism
(Illarionov \& Sunyaev 1975; Chevalier 1989; Lin, Bodenheimer, \&
Woosley 1991; Alpar 2001).

Obviously $r_{\rm A}$ must exceed the neutron star radius (10 km here)
if the field is to have any effect. The above equations thus require a
strong field, $B > 10^{12}$ G, and accretion rates characteristic of
times at least one day.  Additionally there is a critical accretion
rate, for a a given field strength and rotation rate, above which the
co-rotation speed at the Alfven radius will be slower than the
Keplerian orbit speed. In this case the matter will accrete rather
than be ejected. Magnetic braking will thus be inefficient until
$\omega^2 \, r_{\rm A}^3 \, > \, GM$, or
\begin{equation}
\dot M_{26} \ < \ 5.7 \times 10^{-4} \ \omega_3^{7/3} \ \mu_{30}^2,
\end{equation}
where $\omega_3$ is the angular velocity in thousands of radians per
second ($\omega_3$ = 1 implies a period of 2$\pi$ ms). This turns out
to be a very restrictive condition. For a given $\mu_{30}$ and
$\omega_3$, eq. (6) and (8)  give a time, $t_{5, \rm min}$, when
braking can begin.
The torque on the neutron star from that point on will be 
\begin{equation}
I \dot \omega \ = \ \mu^2/r_{\rm A}^3  \ = 10^{60} 
\frac{\mu_{30}^2}{r_{\rm A}^3}, 
\end{equation}
with $I$, the moment of inertia of the neutron star, approximately
10$^{45}$ (Lattimer \& Prakash 2001).
The integrated deceleration will be 
\begin{equation}
\Delta \omega \ = \ 250 \, \mu_{30}^{2/7} \, t_{5, \mathrm
  min}^{-3/7}.
\end{equation}

Putting it all together, a 1.4 $\Msun$ neutron star can be braked to a
much slower speed ($\Delta \omega \sim \omega$) by fall back if
$\mu_{30} >$ 78, 43, or 25 for initial periods of 6, 21, or 60 ms,
respectively. If the surface field strength - for a dipole
configuration is less than $2 \times 10^{13}$ G, braking by the
propeller mechanism will be negligible in most interesting situations.

However, we have so far ignored all non-magnetic forces save gravity
and centrifugal force. A neutron star of age less than one day
is in a very special situation. The accretion rate given by eq. (5) is
vastly super-Eddington. This means that matter at the Alfven radius
will be braked by radiation as well as centrifugal force and the
likelihood of its ejection is greater. Fryer, Benz, \& Herant (1996)
have considered neutron star accretion at the rates relevant here and
find that neutrinos released very near the neutron star actually drive
an {\sl explosion} of the accreting matter. Just how this would all
play out in a multi-dimensional calculation that includes magnetic
braking, rotation, and a declining accretion rate has yet to be
determined, but is obviously a subject worthy of more investigation.

The ejection of material by the propeller mechanism also inhibits the
accretion so the process is self limiting. If this is the way most
neutron stars are slowed, one might expect a correlation of pulsar
period with the mass of the progenitor star. More massive stars
experience more fall back and make slowly rotating neutron
stars. Interestingly the Crab was probably a star of $\sim$10 $\Msun$
and may thus have experienced very little fall back. The matter that
is ejected by the propeller mechanism in more massive stars could
contribute appreciably to the explosion of the supernova and especially
its mixing.

\section{GRBs and Pulsars - Can One Have Them Both?}

We have highlighted an interesting quandary. In order to make GRBs in
the collapsar model one needs a specific angular momentum $j >
10^{16}$ erg s, corresponding to a period of $\ltaprx 0.5$ ms for a
neutron star with moment of inertia $\sim10^{45}$ g cm$^2$.  Other GRB
models need only two or three times less. Yet pulsars are estimated to
have periods at birth in the range 20 - 100 ms, roughly two orders of
magnitude slower. Can reasonable stellar evolution scenarios give
both?

First, we note that the circumstances leading to a GRB are special,
occurring in only $\sim 1$\% of all core collapse supernovae (roughly
1000 GRBs happen in the universe each day; the supernova rate is of
order one per second). The star that dies and makes a GRB must be a
WR-star and was very likely influenced by being in a mass exchanging
binary. Removing the envelope lessens the torque at the edge of the
helium core and allows more rapid rotation at death. But mass loss and
magnetic torques during the WR-phase decrease the angular momentum,
especially if one uses currently favored loss rates. Reducing those
rates, by e.g., $Z_o^{1/2}$ where $Z_o$ is the {\sl initial}
metallicity of the star, certainly goes in the right direction and
would help to explain why GRBs seem to occur more frequently in the
early universe (Stern, Ateia, \& Hurley 2002), but even this does not go far
enough.  The torques given by Spruit (2002) and included in the
calculations of Heger et al. reported at this meeting rule out the
most favored model for GRBs (Table 1). These torques are obviously
uncertain, and perhaps one should be pleased to get within an order of
magnitude of the desired answer, but they do scale as the sixth power
of the shear ($(\frac{r \partial_r \Omega}{\Omega})^6$). Unless this
scaling is itself quite wrong, variations of order factor of 10 in the
overall strength will make little difference in the outcome. It may be
possible to obtain the necessary conditions for GRBs with reasonable
variation of Spruit's parameters, but then isn't one required to use
these same parameters in studies of more common supernovae that make
pulsars?

Factors that would facilitate the production of collapsars or other GRB models
requiring rapid rotation in a massive WR-star would be:

\begin{itemize}

\item{Decreased mass loss} - mass loss removes angular momentum as
well as mass. Does decreasing the initial metallicity appreciably slow
mass loss for WR-stars? 

\item{The angular dependence of mass loss} - if the mass loss is
concentrated at the poles as some models (Maeder \& Meynet 2000)
suggest, less angular momentum is carried away.

\item{Reduced magnetic torques} - Is the formalism of Spruit (2002),
based on the interchange instability, correct for massive stars? Are
composition gradients more effective at inhibiting angular momentum
transport than the default values of parameters in that model would
suggest?

\item{Binary mergers} - Scenarios for keeping the helium core rotating
rapidly by merger with a binary companion (presumably via common
envelope) when the primary is well into helium burning need exploring,
but the frequency of such events must be reasonably high.

\end {itemize}

Perhaps the puzzle is not why GRB progenitors rotate so rapidly as why
pulsars rotate so slowly. Either common neutron stars are born with
long rotational periods or they are braked. We have discussed here
two processes that might brake a young neutron star. Both seem to require
magnetic field strengths substantially in excess of 10$^{13}$
G. Interestingly these braking mechanisms would slow a young
neutron star, but not a black hole and this at least may be seen as
favoring the collapsar model.

This research has been supported by the NSF (AST 02-06111), NASA
(NAG5-12036), and the DOE Program for Scientific Discovery through
Advanced Computing (SciDAC; DE-FC02-01ER41176).  AH is supported, in
part, by the Department of Energy under grant B341495 to the Center for
Astrophysical Thermonuclear Flashes at the University of Chicago, and
a Fermi Fellowship at the University of Chicago.



\begin{references}

\reference
Alpar, M. A. 2001, ApJ, 554, 1245

\reference
Arras, P., Flanagan, E. E., Morsink, S. M., Schenk, A. K., Teukolsky,
S. A., \& Wasserman, I. 2002, submitted to \apj, astro ph-0202345

\reference
Baade, W., \& Zwicky F. 1934, Pub National Acad Sci, 20, 259

\reference
Berger, E., Kulkarni, S. R., \& Chevalier, R. A. 2002, ApJL, 577, 5

\reference 
Chevalier, R. A. 1989, \apj, 346, 847

\reference
Colgate, S. A., \& White, R. H. 1966, \apj, 143, 626

\reference
Duncan, R. C., Shapiro, S. L., \& Wasserman, I. 1986, \apj, 309, 141 

\reference
Frail, D. A., Kulkarni, S. R., Sari, R., Djorgovski, S. G., Bloom,
J. S., Galama, T. J., Reichart, D. E., Berger, E., et al. 2001, ApJL,
562, 55

\reference
Fryer, C. L., Benz, W., \& Herant, M. 1996, \apj, 460, 801

\reference
Fryer, C. L., Woosley, S. E., \& Heger, A. 2001, \apj, 550, 372

\reference
Fowler, W. A., \& Hoyle, F. 1964, \apjs, 9, 201

\reference
Hamaan, W.-R., \& Koesterke, L. 1998, \aap, 335, 1003

\reference
Heger, A., Langer, N., \& Woosley, S. E. 2000, ApJ, 528, 368

\reference
Heger, A., Woosley, S. E., \& Spruit, H. 2003, ApJ, in preparation

\reference
Hoyle, F. 1946, MNRAS, 106, 343

\reference
Illarionov, A. F., \& Sunyaev, R. A. 1979, \aap,39, 185

\reference 
Kaspi, V., \& Helfand, D. 2002, in ASP Conf Series 271,
ed. P. O. Slane \& B. M. Gaensler, San Francisco: ASP, p.3

\reference
Kawabata, K. S., Jeffery, D. J., Iye, M., Ohyama, Y., \& 27 others 2002, 
ApJL, 580, 39

\reference
Lattimer, J. M., \& Prakash, M. 2001, \apj, 550, 426

\reference
LeBlanc, J. M., \& Wilson, J. R. 1970, ApJ, 161, 541

\reference
Li, Z.-Y., \& Chevalier, R. A. 1999, \apj, 526, 716
\reference
Lindblom, L., Tohline, J. E., \& Vallisneri, M. 2001, \prl, 86, 1152

\reference
Lin, D. N. C., Woosley, S. E., \& Bodenheimer, P. H. 1991, Nature,
353, 827

\reference
Maeder, A., \& Meynet, G. 2000, \aap, 361, 159

\reference
MacFadyen, A., Woosley, S. E. 1999 ApJ, 524, 262

\reference
MacFadyen, A., Woosley, S. E., \& Heger, A. 2001, ApJ, 550, 410

\reference
Mestel. L., \& Spruit, H. C. 1987, MNRAS, 226, 57

\reference
Meszaros, P. 2002, ARAA, 40, 137

\reference
Narayan, R., Piran, T.,  \& Kumar, P. 2001, ApJ, 557, 949

\reference
Ostriker, J. P. \& Gunn, J. E. 1971, ApJL, 164, 95O

\reference
Piro, L., Costa, E., Feroci, M., Frontera, F., Amati, L.,
dal Fiume, D., Antonelli, L. A., et al. 1999, ApJL, 514, L73. 

\reference
Piro, L., Garmire, G., Garcia, M., Stratta, G., Costa, E., Feroci, M.,
Meszaros, P., Vietri, M., et al. 2000, Science, 290, 955

\reference
Qian, Y. Z., \& Woosley, S. E. 1996, \apj, 471, 331

\reference
Shapiro, S. L. 2000, ApJ, 544, 397

\reference
Spruit, H.C. 2002 , \aap, 381, 923

\reference
Stern, B. E., Atteia, J.-L., \& Hurley, K. 2002, \apj, 578, 304 

\reference
Thompson, T. 2003, preprint.

\reference
Thompson, T., Burrows, A., \& Meyer, B. 2001, 562, 887

\reference
Usov, V. 1992, Nature, 357, 472

\reference
Vietri, M., \& Stella, L. 1998, ApJL, 507, L45 

\reference
Vietri, M., \& Stella, L. 1999, ApJL, 527, L43 

\reference
Wheeler, J. C., Yi, I., H\"oflich, P., \& Wang, L. 2000, ApJ, 
537, 810

\reference
Wheeler, J. C., Meier, D. L., \& Wilson, J. R. 2002, ApJ, 568, 807

\reference
Wellstein, S., \& Langer, N. 1999, \aap, 350, 148

\reference
Woosley, S. E. 1993, ApJ, 405, 273.


\end{references}
\end{document}